\documentclass[12pt,preprint]{aastex}

\usepackage[hyphens]{url}

\shorttitle{Gamma Ray Signal from \object{PSR~B1259$-$63/LS~2883}}
\shortauthors{Khangulyan et al.}

\begin{document}


\title{Gamma Ray Signal from the Pulsar Wind in the Binary Pulsar system \object{PSR~B1259$-$63/LS~2883}}


\author{ Dmitry Khangulyan\altaffilmark{1},  Felix
  A. Aharonian\altaffilmark{2,3}  Sergey
 V. Bogovalov\altaffilmark{4},  Marc Rib\'o\altaffilmark{5}  \vspace{5mm}}

\affil{$^{1}$Institute of Space and Astronautical Science/JAXA, 3-1-1
  Yoshinodai, Chuo-ku, Sagamihara, Kanagawa 252-5210,
  Japan\email{khangul@astro.isas.jaxa.jp}}
%
\affil{$^{2}$Dublin Institute for Advanced Studies, 31 Fitzwilliam
  Place, Dublin 2, Ireland\email{felix.aharonian@dias.ie}}
\affil{$^{3}$Max-Planck-Institut f\"ur Kernphysik, Saupfercheckweg 1,
  D-69117 Heidelberg, Germany}  
%
\affil{$^{4}$National research nuclear university-MEPHI, Kashirskoe
  shosse 31, Moscow, 115409 Russia\email{svbogovalov@mephi.ru}}
\affil{$^{5}$Departament d'Astronomia i Meteorologia, Institut de
  Ci\`ences del Cosmos (ICC), Universitat de Barcelona (IEEC-UB),
  Mart\'{\i} i Franqu\`es 1, E-08028 Barcelona,
  Spain\email{mribo@am.ub.es}}


\begin{abstract}
  Binary pulsar systems emit potentially detectable components of gamma ray emission
  due to Comptonization of the optical radiation of the
  companion star by relativistic electrons of the pulsar wind, both
  before and after termination of the wind. The recent optical
  observations of binary pulsar system \object{PSR~B1259$-$63/LS~2883}
  revealed radiation properties of the companion star which differ
  significantly from previous measurements.  In this paper we study
  the implications of these observations for the interaction rate
  of the unshocked pulsar wind with the stellar photons and the related 
  consequences for fluxes of high energy (HE) and very high
  energy (VHE) gamma rays.  We show that the signal should be strong
  enough to be detected with {\it Fermi} close to the periastron
  passage, unless the pulsar wind is strongly anisotropic or the
  Lorentz factor of the wind is smaller than $10^3$ or larger that
  $10^5$.  The higher luminosity of the optical star also has two
  important implications: (i)  attenuation of  gamma rays 
  due to  photon-photon  pair production, and (ii) Compton drag of the unshocked wind.  
 While the first effect has an  impact on the lightcurve  of VHE gamma rays, 
 the second  effect may significantly decrease the energy available for  
 particle acceleration  after termination of the wind. 
\end{abstract}


\keywords{
binaries: close ---
gamma rays: stars --- 
pulsars: individual (PSR~B1259$-$63)
}


\section{Introduction} \label{sec:intro} 

Three binary systems containing a massive star and a compact object --
\object{LS~5039}, \object{LS~I~+61~303} and \object{PSR~B1259$-$63}--
have been clearly detected in TeV energy band (see
\url{http://tevcat.uchicago.edu/} for the updated information). While
the nature of the compact companion in \object{LS~5039} and
\object{LS~I~+61~303} is not yet established
\citep{casares05a,casares05b,sarty11}, the detection of the pulsed radio
emission from \object{PSR~B1259$-$63} indicates the presence of a
47.7~ms pulsar in the system \citep{johnston92}. The pulsar orbits a
luminous star in a very eccentric orbit with the following orbital
parameters: eccentricity $e=0.87$, period $P_{{\rm orb}}= 1237$~d, and
semi-major axis $a_{\rm 2}=6.9\rm \, AU$ \citep[see][and references
therein]{negueruela10}. The system displays variable broadband
nonthermal radio, X-ray and TeV gamma ray emission close to periastron
passage \citep{johnston05,uchiyama09,masha09,grove95,aharonian05,aharonian09},
which currently lacks successful multiwavelength
interpretations. Moreover, the {\it Fermi} LAT observations of periastron
passage in December 2010 have shown that in general the GeV flux level
from the system is quite low, although a short intensive flare was
detected as well \citep[see e.g.][]{tam11,abdo11}.

Recently, optical observations with \facility{VLT UT2 Kueyen}
discovered that the optical star \object{LS~2883} corresponds to a late
O-star and has a significantly higher luminosity of $L_*=2.3\times10^{38}\rm
\, erg\,s^{-1}$ than previously thought \citep{negueruela10}. Because
of fast rotation the star is significantly oblated with equatorial
radius of $R_{\rm eq}=9.7 R_\sun$ and the polar radius of $R_{\rm
  pole}=8.1 R_\sun$. This leads as well to a strong gradient of the
star surface temperature with $T_{\rm eq}=27\,500$~K and $T_{\rm
  pole}=34\,000$~K. The star rotation axis is inclined by
$i_*\simeq33^\circ$ in respect to the line-of-sight
\citep{negueruela10}.
The distance to the system is now estimated to be $2.3\pm0.4\,\rm
kpc$. Moreover, the observations favored an orbital inclination value
of $i\simeq25^\circ$, which is remarkably smaller than the previously
obtained value of $\sim35^\circ$\citep{johnston94}. All these new
parameters together should have an important impact on the
multiwavelength properties of this system.

The orbital separation distance, pulsar spindown luminosity and the
lack of the accretion features suggest a realization of the
compactified nebula scenario, i.e. the source contains two distinct
regions: the relativistic pulsar wind and the terminated flow
\citep{tavani97,bogovalov08}. The VHE emission is expected to
originate in the post termination shock region, and a number of models
have been proposed invoking both hadronic \citep{kawachi04,neronov07}
and leptonic \citep{kirk99,khangulyan07} radiation mechanisms. In the
framework of the hadronic scenario, the two humped TeV lightcurve
obtained with HESS in 2004 \citep{aharonian05} have been interpreted
as the enhancement of the production rate due to the pulsar passage
through the dense stellar disc.  However, the recent report of a more
complicated TeV gamma ray lightcurve by HESS with multiple humps and
deeps disfavors, to a large extent, the hadronic scenario
\citep{aharonian09}. In the case of leptonic origin of the emission,
there is a number of additional effects, which may significantly
affect the production rate of the nonthermal emission in the post
shock region \citep{khangulyan07}. Importantly, in the leptonic
scenario, one expects a specific radiation component from the
unshocked pulsar wind.  Although, the pulsar wind region is not
expected to produce VHE emission, a line-like bulk Comptonization
component from this region in HE band is predicted for isolated
pulsars like Crab pulsar \citep{bogovalov00} and for the binary pulsar
system \object{PSR~B1259$-$63} \citep{ball00,ball01,khangulyan07}.  It
is difficult to overestimate the importance of observational proof of
this radiation component; it offers a unique opportunity of
detecting a direct signal from the pulsar wind. The most meaningful
constraints in this regard can be obtained in the 1 to 100 GeV
region. Therefore the observations with {\it Fermi} and {\it AGILE} gamma ray
telescopes close to the periastron passage are of great interest; any
result (flux upper limit or detections of a signal) can
greatly contribute to our understanding of the physics of  pulsar winds.

Since the interaction of the pulsar wind with the stellar photon field
does not occur in the {\it saturation regime}, the production rate is
very sensitive to the properties of the photon field. In this paper we
present new calculations of the radiation signal from the pulsar wind
taking into account the new properties of the optical star \citep{negueruela10} and using
results of detailed hydrodynamical modeling of the interaction between the
pulsar and stellar winds in \object{PSR~B1259$-$63/LS~2883} \citep{bogovalov08}.

\section{Pulsar Wind in Binary Pulsar System}

In the framework of the generally accepted paradigm \citep{kennel84},
pulsars launch cold ultrarelativitic winds which are terminated due to
external pressure. At the termination shock, the wind electrons can be
accelerated to multi-TeV energies. The radiation of these electrons
results in a phenomenon called pulsar wind nebula. Since the wind is
cold, i.e. particles remain at rest in the wind co-moving system, no
synchrotron emission is expected from the wind before its
termination. On the other hand, the comptonization of the
ultrarelativistic wind by external radiation fields, through which the
wind propagates, can lead to detectable gamma ray emission. This
effect is relatively weak in isolated pulsars, and can achieve a
reasonable efficiency only in powerful pulsars, provided that the
particle dominated wind is formed close to the light cylinder
\citep{bogovalov00}.  In binary systems, the process operates with an
enhanced efficiency thanks to the presence of the dense radiation
field of the optical companion \citep{ball00,ball01}. The interaction
rate in this channel depends on different parameters characterizing
the system: (i) luminosity and temperature of the optical star; (ii)
orbital separation and inclination; (iii) distance to the system; (iv)
size of the region occupied by the pulsar wind; (v) pulsar wind bulk
Lorentz factor.

Remarkably, the recent optical observations of \object{PSR~B1259$-$63}
have significantly revised the parameters (i), (ii) and (iii) in favor
of higher temperature and luminosity, smaller inclination angle and
further system location.  Regarding the wind bulk Lorentz factor, it
remains a highly uncertain parameter.

The size of the region, where the pulsar wind can propagate depends on
the ratio of the wind ram pressures, $\eta$ \citep{bogovalov08}. Since the
pulsar spindown luminosity is known, one can estimate the ram pressure
of the pulsar wind
\begin{equation}
  P_{\rm pw}={L_{\rm sd}\over 4\pi r^2 c}\,,
\end{equation}
where $L_{\rm sd}$ and $r$ are the spindown luminosity of the pulsar
and distance to the pulsar, respectively. It should be noted that this
relation ignores the possible effects related to the anisotropy of the
pulsar wind. Although, the level of the anisotropy may be quite high
at large distances from the pulsar, e.g. in the case of the Crab-like
pulsars \citep{bogovalov02}, in this paper we limit our consideration
by the case of an isotropic wind. 

To obtain the ram pressure of the
stellar wind, one needs detailed information about the properties of
the optical star, including the mass-loss rate and wind velocity
profile, which are currently not firmly
established. For the given optical star luminosity, the mass-loss
rate can be estimated at the level of $\dot{M}=6\times10^{-8} M_\sun \rm yr^{-1}$
\citep{vink00}. Accounting for the wind velocity
at interaction point $V_{\rm w}<V_\infty=1350\pm200\,\rm km\, s^{-1}$
\citep{mccollum93}, it is possible to estimate the expected value of
the ratio of the momentum flux densities, $\eta$-parameter, as follows:
\begin{equation}
\eta={L_{\rm sd}\over \dot{M}c V_{\rm w}}=5\times10^{-2}\left({\dot{M}\over6\times10^{-8}M_\sun\rm yr^{-1}}\right)^{-1}\left({V_{\rm w}\over 1350\,\rm km\,s^{-1}}\right)^{-1}\,.
\label{eq:eta}
\end{equation}
We should note, however, that there are  several factors  which  may introduce 
significant uncertainties in the $\eta$-parameter. In particular, the wind porosity may lead to an
overestimation of the mass-loss rate of the star \citep{owocki06} and
consequently to a significant underestimate of the $\eta$ parameter
value. The opposite situation may occur if the pulsar wind interacts with
the stellar wind close to the star equatorial plane, where a dense
Keplerian disk is formed. Since the disk is expected to have a
significantly higher density than the polar wind, and its typical velocity at distance $r$
(i.e. Keplerian velocity) may be as high as
\begin{equation}
v_{\rm disk}\simeq 200 \left({r\over 10^{13}\rm cm}\right)^{-1/2}\rm km\,s^{-1}\,,
\end{equation}
the disk effective ram pressure may significantly exceed the polar
wind one, i.e. the $\eta$-parameter may be remarkably smaller than the
estimate of Eq.(\ref{eq:eta}).  Moreover, because of the disk rotation
and pulsar orbital velocity, the structure of the wind termination
shock, in respect to the observer direction, may be rather different
for two pulsar--disk interaction points. Because of these uncertainties
related to the value of the $\eta$-parameter, below we will consider a
fairly broad range of the $\eta$ parameter.

\begin{figure}
\resizebox{\columnwidth}{!}{
\includegraphics[angle=270,clip]{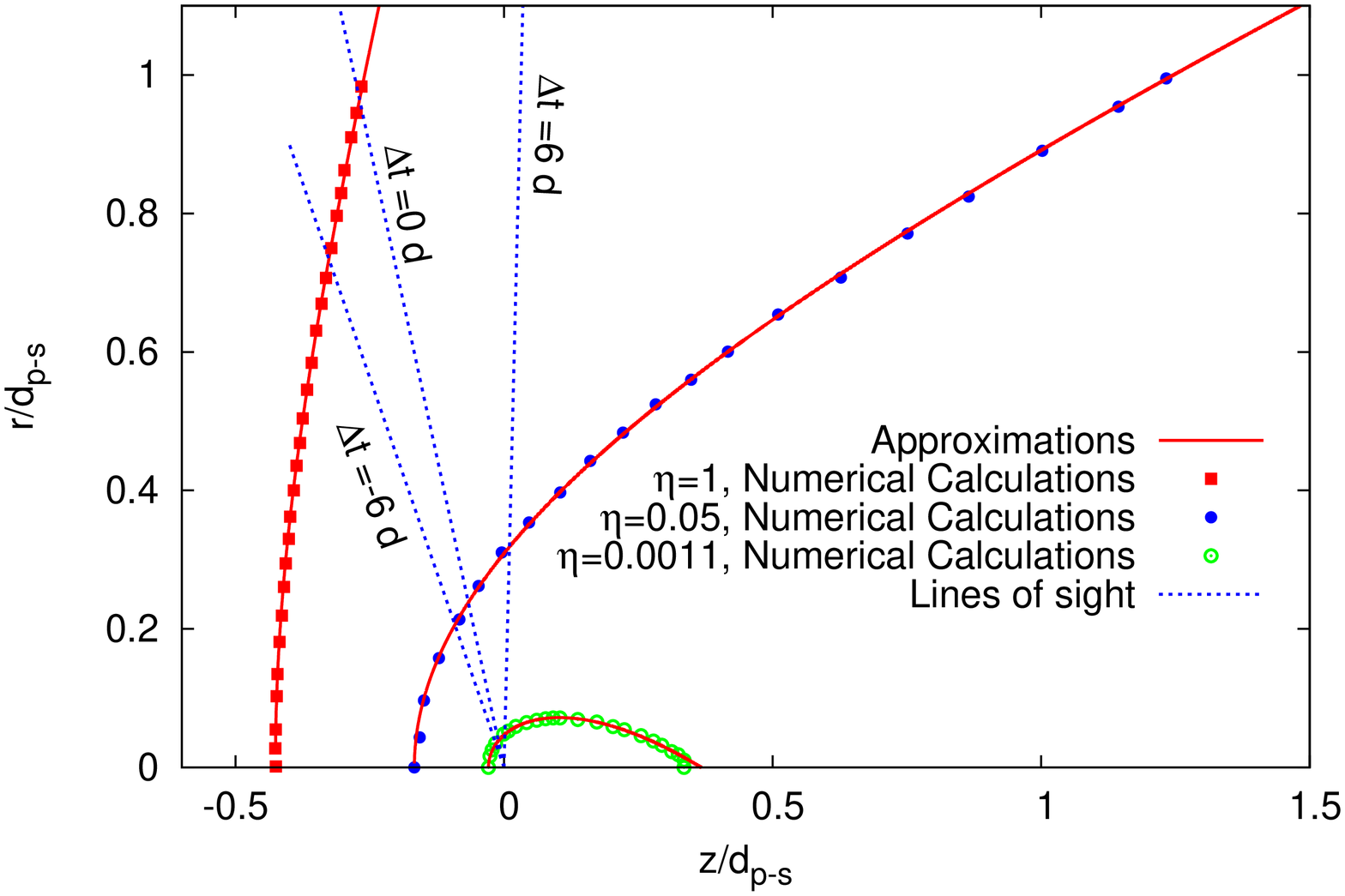}}
\caption{The geometry of interaction of  pulsar and optical star winds: the pulsar is assumed to be located at the point with coordinates $r=0,\,z=0$, the optical star at $r=0,\,z/d_{\rm p-s}=-1$.  The shapes of the termination shocks as obtained through the numerical modeling \citep{bogovalov08} are shown for the following values of the $\eta$-parameter: $\eta=1$ (squares), $\eta=0.05$ (filled circles) and $\eta=1.1\times10^{-3}$ (open circles). The analytical approximations Eq.(\ref{eq:ap1}-\ref{eq:ap11}) are shown with solid lines. The directions towards the observer are shown with dotted lines for three different epochs: $-6$, $0$ and $6$ days to periastron passage.  }
\label{fig:shock}
\end{figure}

In Figure \ref{fig:shock} the shapes of the termination shock for three
different values of the $\eta$ parameter are shown. The points in the
figure are from the results of numerical modeling performed by
\citet{bogovalov08}, for $\eta=1$ (squares), $\eta=0.05$ (filled
circles) and $\eta=1.1\times10^{-3}$ (open circles).  Here the value of
$\eta=1$ roughly corresponds to the case of the interaction with the
clumpy polar wind; $\eta=0.05$ to the case of collision with the
stellar wind; and $\eta=1.1\times10^{-3}$ is a lower limit value,
which can be realized if e.g. pulsar wind is significantly anisotropic
at binary system scales; or if the stellar disk plays an important
role in the interaction. To simplify the calculations we have
approximated the termination shock by the following
analytical expressions: \\
for $\eta=1$
\begin{equation}
r=3.02\sqrt{(z+0.425649)(z+0.92)}\,,
\label{eq:ap1}
\end{equation}
for $\eta=0.05$
\begin{equation}
r= 0.33\sqrt{(z+0.167)(z+5.25)}\,,
\label{eq:ap05}
\end{equation}
for $\eta=1.1\times10^{-3}$
\begin{equation}
r=0.75\sqrt{(z+0.0289564)}(0.367-z) \, .
\label{eq:ap11}
\end{equation}
Here $z$ is the coordinate along the axis joining the 
pulsar  and the star (it is assumed that the pulsar is located at the point ``0'',
and the optical star is located at $(z=-1,\, r=0)$), and $r$ is
the corresponding cylindrical radius (both coordinates are
measured in the pulsar-star separation units, $d_{\rm p-s}$).  Eq.(\ref{eq:ap11}) shows
non-smooth behavior at $z\simeq0.367$. It does not have a physical
meaning, but is the result of the simplification of the procedure
which neglects the complex structure of the termination shock wave
close to the symmetry axis \citep[see][for details]{bogovalov08}.

 Since the particles in the pulsar wind move radially, the emission
towards the observer is produced by electrons which were moving in
this direction initially.  The expected gamma ray flux is calculated
as
\begin{equation} {{\rm d}N\over{\rm d}E_\gamma{\rm d}S{\rm
      d}t}={c\over 4\pi d^2}\int\limits_{\rm pulsar}^{\rm
    termination\,\,shock}{\rm d}l\,\int{\rm d}\epsilon_{\rm ph}\, \int
  {\rm d} E_{\rm e} \left(1-\cos\theta\right){\rm e}^{-\tau}{{\rm d}
    \sigma\over{\rm d}E_\gamma}{{\rm d}N_{\rm e}\over {\rm d}E_{\rm
      e}{\rm d}l}{{\rm d}N_{\rm ph}\over {\rm d}\epsilon_{\rm ph}{\rm
      d}V}\,.
\label{eq:flux}
\end{equation}
Here $d$ is the distance to the system; ${\rm d} \sigma/{\rm
  d}E_\gamma$ is the differential anisotropic inverse Compton
cross-section \citep{aharonian81,bogovalov00}; $\tau$ is the energy
dependent optical depth due to gamma-gamma attenuation from the
gamma ray creation point to the observer; and ${\rm d}N_{\rm ph}/ {\rm
  d}\epsilon_{\rm ph}{\rm d}V$ is the target photon density at the
given location. The term representing the electron density in the cold
pulsar wind has the following form:
\begin{equation} {{\rm d}N_{\rm e}\over {\rm d}E_{\rm e}{\rm
      d}l}={L_{\rm sd}\over \Gamma_0mc^3}\delta\left(E_{\rm
      e}-\Gamma_0mc^2-\int\limits_0^l{\rm d}l'\dot{E}/c\right)\,,
\label{eq:electrons}
\end{equation}
where $\Gamma_0$ and $\dot{E}$ are the initial wind bulk Lorentz
factor and the electron energy loss rate.

The integration of Eq.(\ref{eq:flux}) is performed over the line of
sight from the pulsar location to the pulsar wind termination
shock. Obviously, the integration path depends strongly on the orbital
phase. In Figure~\ref{fig:shock}, the lines of sight for three different
orbital phases ($-6$, $0$ and $6$ day from periastron passage) are shown by
dashed lines. 

Another effect, which may lead to an additional orbital phase
dependence, is the shape of the optical star and temperature change
between different regions of the star. To study this effect we
performed calculation for the precise properties of the star,
i.e. assuming the star to be an oblate spheroid with a linear gradient
of the surface temperature as a function of the zenith angel. The
orientation of the star in respect to the observer is defined by
inclination angle, i.e. angle between the star rotation axis and
line-of-sight, which was assumed to be $i_*=33^\circ$, as inferred by
\citet{negueruela10}. To fix the star orientation an additional angle is
required, namely the angle which describes the turn in the plane of the
sky. This angle was assumed to be a free parameter, and its influence
was studied. Numerical calculations show (see Figures~\ref{fig:per} and
\ref{fig:before}) that independently on this parameter, the emission
is well described by a model with a spherical star of radius
$R_*=6.2\times10^{11}\rm cm$ and surface temperature
$T_*=3\times10^4\rm K$. Given the uncertainties related to the
orientation of the star, in what follows we perform calculations for
the spherical star with the inferred parameters.

In Figure~\ref{fig:per} we show the spectral energy distributions
(SEDs) expected at the orbital phase corresponding to the periastron
passage for $\eta=1$ (solid lines); $\eta=0.05$ (dotted lines);
$\eta=1.1\times10^{-3}$ (dashed line). In calculations we have adapted
the following values of the initial wind Lorentz factors:
$\Gamma_0=10^4, 4.6\times10^4, 2.2\times10^5$ and $10^6$. In
Figure~\ref{fig:before} we show a similar plot, but for the orbital
phase corresponding to the epoch of $30$ days before the periastron
passage. For a rather broad range of the wind bulk Lorentz factors
around $\Gamma_0=10^4$, the obtained flux level is above the {\it
  Fermi} sensitivity level (unless the $\eta$-parameter is small
$\eta\ll0.05$).

\begin{figure}
\resizebox{\columnwidth}{!}{
\includegraphics[angle=270,clip]{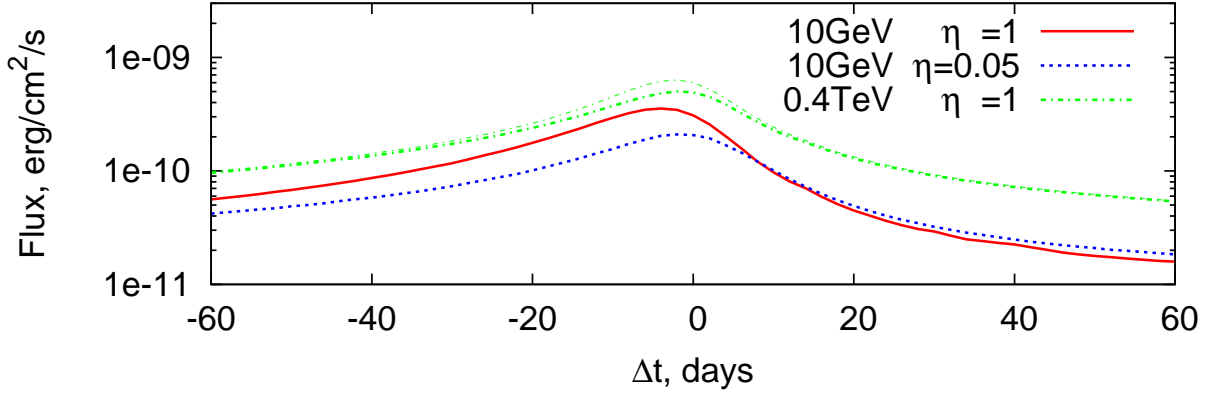}}\\
\caption{ The solid and dotted lines show light-curves
  of 10GeV emission for two different values of the $\eta$-parameter:
  $\eta=1$ (solid lines), $\eta=0.05$ (dotted lines) for the initial
  pulsar wind bulk Lorentz factor of $\Gamma_0=4.6\times10^4$. Light
  curve of 0.4TeV gamma rays, calculated for $\eta=1$ and the initial
  bulk Lorentz factor of $\Gamma_0=10^6$, is shown with thick
  dash-dotted line (accounting for gamma-gamma absorption) and with
  thin dash-dotted line without gamma-gamma attenuation. The
  calculations are performed for a spherical star with radius
  $R_*=6.2\times10^{11}\rm cm$ and surface temperature
  $T_*=3\times10^4\rm K$. }
\label{fig:lc}
\end{figure}

Due to the orbital phase dependence of several key parameters, in
particular, separation distance, the location of the termination
shock, the gamma-gamma optical depth, and the electron -- target
photon interaction angle, the pulsar wind signal has an orbital
phase-dependence. Since the lightcurves for different energies are
quite similar, we show just a few examples. In particular,  the lightcurve for
10~GeV gamma rays is shown in Figure~\ref{fig:lc}. Here we assumed for
the initial pulsar wind Lorentz factor $\Gamma_0=4.6\times10^4$.  In
calculations we used two different values of the $\eta$-parameter:
$\eta=1$ (solid line) and $\eta=0.05$ (dotted line). We note that this
parameter may affect not only the flux level, but also the location of
the light-curve maximum: in the case of the small value the maximum is
located close to the periastron passage, while in the case of larger
values of $\eta$ the maximum is located a few days before periastron
passage.

In the case of large bulk Lorentz factor, $\Gamma_0\sim10^{6-7}$, the
inverse Compton (IC) signal from the pulsar wind may appear at energies beyond the range
of {\it Fermi}/LAT.  In this energy band, the atmospheric Cherenkov
telescope arrays are more appropriate tools for probing the wind's
Lorentz factor (note that for this specific source, currently only the
HESS array is able to monitor \object{PSR~B1259$-$63}). In
Figure~\ref{fig:lc} we show a light curve of 0.4~TeV gamma rays
calculated for the $\eta$-parameter $\eta=1$ and for the bulk Lorentz
factor of $\Gamma_0=10^6$.

\section{Impact of the higher  luminosity of the optical star}
\begin{figure}
\resizebox{\columnwidth}{!}{
\includegraphics[angle=270,clip]{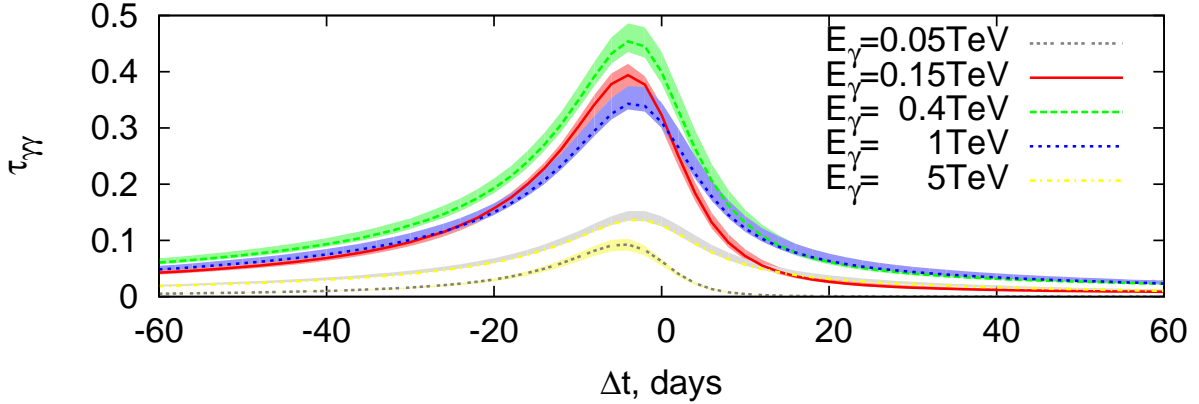}}
\caption{The optical depth for gamma-gamma attenuation from the
  location of the pulsar to the observer for different energies of the
  gamma rays.  The lines show model calculations performed for a
  spherical star with radius $R_*=6.2\times10^{11}\rm cm$ and surface
  temperature $T_*=3\times10^4\rm K$; the filled regions correspond to
  calculations with oblate star for possible star orientations. }
\label{fig:tau}
\end{figure}

In addition to gamma-radiation of the unshocked pulsar wind, we expect
gamma rays (at higher energies) from the Compton scattering of
shock-accelerated electrons \citep{kennel84}.  If the optical radiation
density exceeds the density of the magnetic field, the IC gamma ray
production proceeds in the saturation regime, thus the increased
luminosity of the optical star does not lead to amplification of the
VHE gamma ray signal. On the other hand, for the recently reported
luminosity of the optical star \citep{negueruela10}, the gamma-gamma
opacity for VHE photons traveling from the pulsar to the observer, may
be as large as $0.5$, thus emission may be attenuated by a factor of
1.6.  In Figure~\ref{fig:tau} we show the corresponding optical depth as a
function of the phase for four different energies of gamma rays:
$E_\gamma=0.05$, $0.15$, $0.4$, $1$ and $5$~TeV. Note that the gamma ray
absorption is strongest at the energy of 0.4~TeV, while at energies
below 100 GeV and at multi-TeV energies it becomes negligible. We note
however that the actual absorption level depends on the production
region location, while the calculations in Figure~\ref{fig:tau} assume
that the production occurs in the pulsar location. In particular, in
Figure~\ref{fig:lc} two lightcurves for $0.4$~TeV gamma rays are
shown. The thick dash-dotted line corresponds to the flux level
corrected for the gamma-gamma attenuation, while thin dash-dotted line
shows the intrinsic flux level. It can be seen that the 
attenuation is somewhat weaker than it is expected from the opacity
shown in Figure~\ref{fig:tau}. The reason for that is rather simple,
namely since the gamma ray production occurs along the line of sight
in the pulsar wind zone (see Eq.(\ref{eq:flux})), some gamma ray
photons suffer a weaker attenuation than the one shown in
Figure~\ref{fig:tau}.

\begin{figure}
\resizebox{\columnwidth}{!}{
\includegraphics[angle=270,clip]{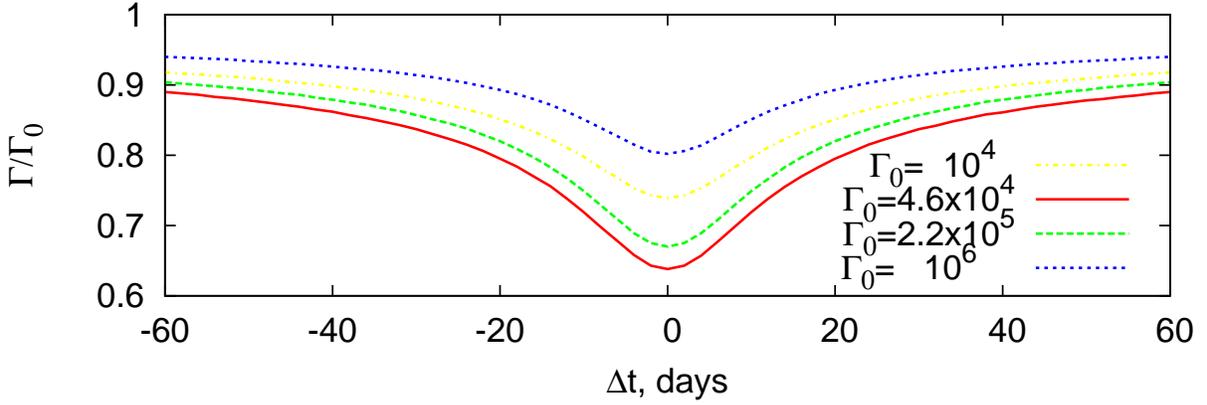}}
\caption{The averaged ratio of the electron Lorentz factor $\Gamma$ at
  the termination shock to the initial value $\Gamma_0$. The averaging
  is performed for electrons propagating within the angle of
  $60^\circ$ towards the optical star. The ratio is calculated for the
  different values of the initial pulsar wind bulk Lorentz factor:
  $\Gamma_0=10^4$ (dash-dotted line), $4.6\times10^4$ (solid line),
  $2.2\times10^5$(dashed line) and $10^6$ (dotted line).  The
  $\eta$-parameter was assumed to be $\eta=1$, and the calculations
  are performed for a spherical star with radius
  $R_*=6.2\times10^{11}\rm cm$ and surface temperature
  $T_*=3\times10^4\rm K$.}
\label{fig:drag} 
\end{figure}

Another important implication of the enhanced optical luminosity of
the companion star for the production of VHE gamma rays is related to
the so-called effect of Compton drag which leads to the reduction of
the Lorentz factor of the wind before it terminates.
\citet{khangulyan07} have argued that this effect could lead to a
deficit of power available for acceleration of VHE particle at the
termination shock.  It has been shown, in particular, that if the
optical star would have a luminosity close to $4\times10^{38}\rm
erg\,s^{-1}$, then the tendency of reduction of the TeV flux observed
by HESS \citep{aharonian05} close to the periastron passage may be
explained by the reduction of the overall energy of the pulsar wind
due to the Compton drag.  The stellar luminosity reported by
\citet{negueruela10} is remarkably close to the one speculated in
\citet{khangulyan07}, thus this effect now becomes more relevant to
the observed TeV lightcurve.  To describe the effect quantitatively,
we have calculated the energy fraction lost by electrons emitted
within the solid angle of $\pi$ steradian towards the optical star.
The result are shown in Figure~\ref{fig:drag} for different values of
the initial wind Lorentz factor: $\Gamma_0=10^4$ (dash-dotted line),
$4.6\times10^4$ (solid line), $2.2\times10^5$ (dashed line) and $10^6$
(dotted lines). The $\eta$-parameter was assumed to be $\eta=1$. The
calculations show that this effect may lead to a significant decrease
of the pulsar wind energy transported to the termination shock in the
case of high values of the $\eta$-parameter, i.e. $\eta\ge0.1$. This
should lead to a proportional decrease of the VHE gamma ray
production. Obviously, the Compton drag can be important only if the
GeV flux from the pulsar wind is high. This radiation should be
clearly detected by {\it Fermi}/LAT, unless the pulsar wind is
strongly anisotropic. The detection of the pulsar wind signal with {\it
  Fermi} could allow a quantitative estimate of the expected decrease
of the VHE gamma ray production.

The Compton drag should lead as well to a decrease of the pulsar wind
ram pressure in the collision region. However, given a relatively weak
dependence of the termination shock shape on the $\eta$-parameter, it
is unlikely that the Compton drag would change significantly the
pulsar~--~optical star interaction regime.

\section{Conclusion}
\begin{figure}
\resizebox{\columnwidth}{!}{
\includegraphics[angle=270,clip]{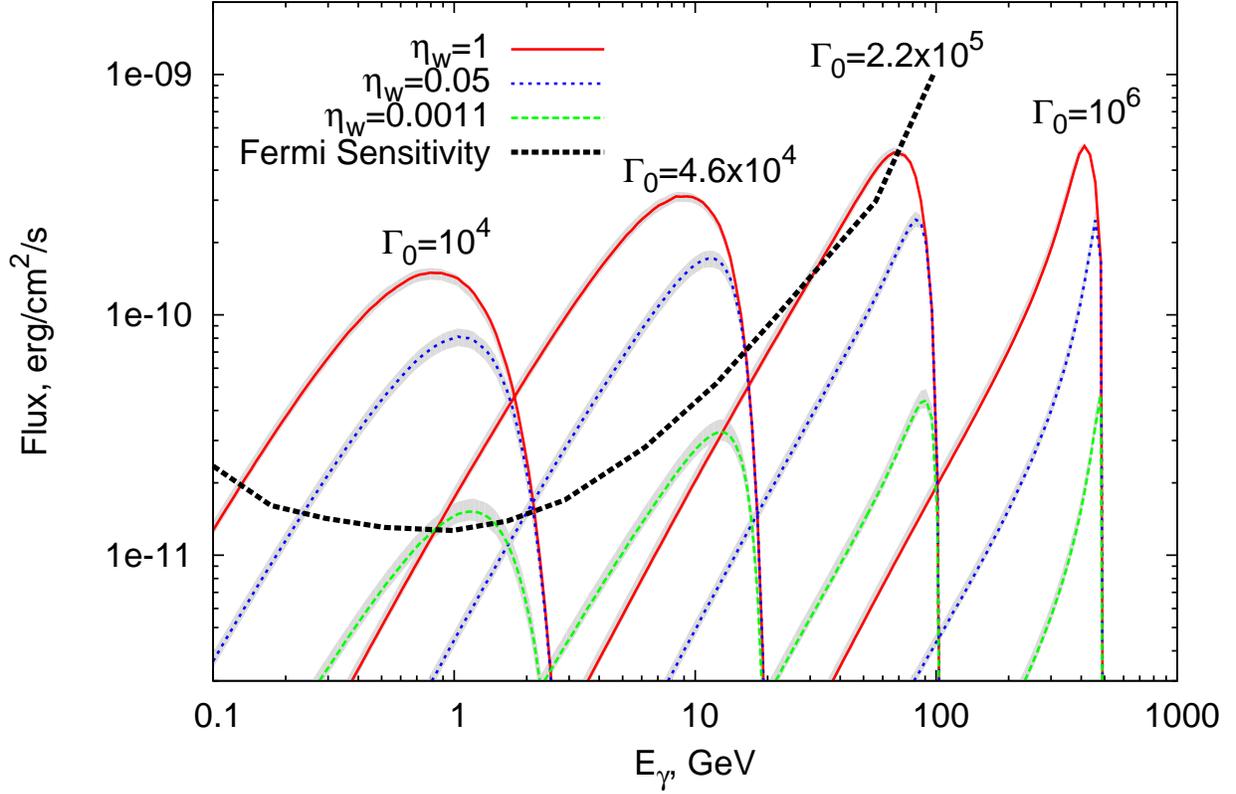}}
\caption{Spectral energy distributions of IC radiation from the unshocked pulsar
  wind for the epoch of periastron passage. The calculations were
  performed for different values of the $\eta$-parameter: $\eta=1$
  (solid lines), $\eta=0.05$ (dotted lines) and
  $\eta=1.1\times10^{-3}$ (dashed lines). The initial pulsar wind bulk
  Lorentz factor was assumed to be $\Gamma_0=10^4$, $4.6\times10^4$,
  $2.2\times10^5$ and $10^6$. The thick dashed line roughly
  corresponds to the expected {\it Fermi} sensitivity for $0.1\rm yr$
  observation. The lines show model calculations performed for a
  spherical star with radius $R_*=6.2\times10^{11}\rm cm$ and surface
  temperature $T_*=3\times10^4\rm K$; the regions filled with gray
  correspond to calculations with oblate star for possible star orientations.}
\label{fig:per}
\end{figure}

Motivated by the recent revision of optical properties of LS~2883, the
companion star of \object{PSR~B1259$-$63} \citep{negueruela10}, we
present new calculations of high energy gamma ray fluxes using a
realistic termination shock geometry as described in
\cite{bogovalov08}. Calculations show that the higher optical star
luminosity is compensated, to a large extent, by the new estimate of
the distance to the source. Thus, the new calculations of gamma ray
fluxes are quite close to previous predictions based on the old gamma
ray parameters of the optical star \citep[see in][]{khangulyan07}.
According to Figure~\ref{fig:per}, the 0.1--10 GeV gamma ray fluxes
calculated for epochs close to the periastron passage are below the
current upper limits obtained with EGRET \citep{tavani96}, and above
the minimum fluxes detectable by {\it Fermi}/LAT for observation time
of about 1 month.  The pulsar wind radiation component can be
identified by its distinct spectral shape. Another important feature
of this emission is the expected modulation with the pulsar period on
the top of the smooth orbital phase dependence. Indeed, since the
emitting electrons move towards the observer almost with the speed of
light, the gamma ray signal should have the second modulation
reflecting the time structure of the striped pulsar wind. Although, a
detailed shape of the fine lightcurve can be hardly obtained given a
lack of any consistent description of the striped pulsar wind, a
detailed search for this effect looks quite important for a
consistent interpretation of the results obtained with {\it Fermi}/LAT
\cite{tam11,abdo11}.

\begin{figure}
\resizebox{\columnwidth}{!}{
\includegraphics[angle=270,clip]{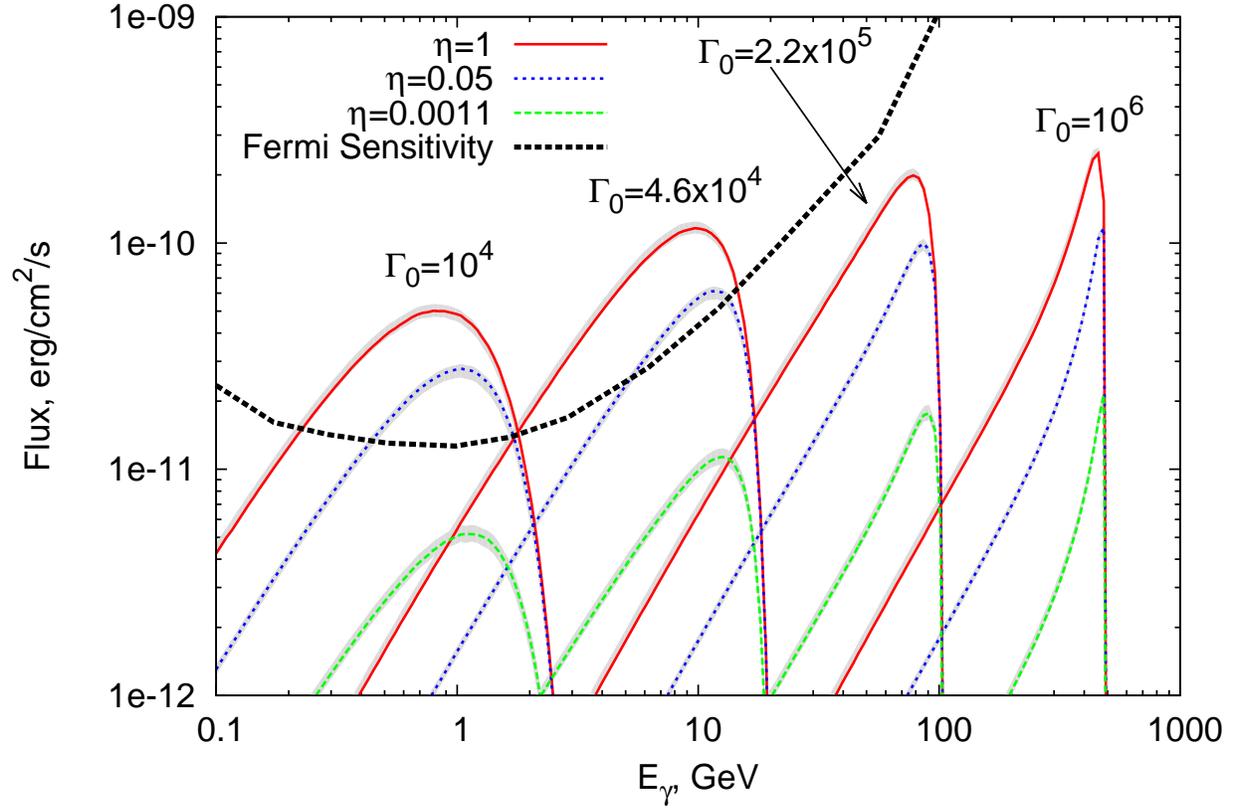}}
\caption{The same as in Fig.\ref{fig:per} but for the epoch of 30 days before periastron passage.}
\label{fig:before}
\end{figure}

The most important implication of detection of this component of
gamma radiation would be the unique opportunity to measure the
Lorentz factor of the pulsar wind.  On the other hand, in the case of
failure of detection of this component at GeV and/or TeV energies, the conclusions could be
equally interesting and important. The possible reasons of
non-detection of gamma rays from the unshocked wind could be: (i) an
extremely powerful stellar wind, i.e. very low values of the $\eta$
parameter; (ii) unconventional, i.e. very low ($\Gamma_0 \ll 10^4$) or
very large ($\Gamma_0 \geq 10^6$) values of the pulsar wind bulk
Lorentz factor.  The first condition requires the pulsar to interact
with the stellar disc all over the orbit. This implies a very
specific realization in the sense of orientation of the stellar disc
(namely the orbital plane and the disc plane should almost coincide),
which contradicts to the current expectations
\citep{melatos95,bogomazov05,bogovalov08,kerschhaggl10}. However, we should note
that one cannot exclude that the pulsar wind is strongly anisotropic.
If so, the gamma ray signal should be anisotropic as well. This can be
another reason for reduction of the gamma ray flux, which
unfortunately would make the conclusions concerning the range of
parameters $\Gamma_0$ and $\eta$ less robust.  Finally, one should
mention that if the pulsar wind is not absolutely cold, electrons
in the frame of the wind might have a rather broader distribution. This
would make the gamma ray spectrum less distinct and smoother compared
to the ones shown in Figures~\ref{fig:per} and \ref{fig:before}.
 
Regarding  VHE energy gamma rays produced after termination of the wind, 
the new optical observations of \citet{negueruela10}  imply a significant reduction 
of the flux of IC gamma rays  produced by shock-accelerated electrons. 
 All three main  factors   related to (1) the larger distance to the 
source, (2) the gamma-gamma attenuation, and (3) the Compton drag of the pulsar 
wind work in the same (negative) direction reducing the gamma ray flux by 
a factor of up to 10.  Given that  the previous studies based on the old optical observations 
already have required  a significant fraction of the spin-down luminosity (5\%--10\%) to be released in 
TeV gamma rays,   the revised energy requirements become almost 
unbearable. A possible solution to the energy budget crisis could be  the Doppler 
boosting of radiation as suggested in  \citet{khangulyan08}.  
This important issues  will be discussed elsewhere. 

\acknowledgments 
The work of S.V.Bogovalov have been supported by the
Federal Targeted Program "The Scientific and Pedagogical Personnel of
the Innovative Russia" in 2009-2013 (the state contract N 536 on May
17, 2010). M.R. acknowledges support by the Spanish Ministerio de Ciencia e 
Innovaci\'on (MICINN) under grant FPA2010-22056-C06-02, as well as financial 
support from MICINN and European Social Funds through a \emph{Ram\'on y 
Cajal} fellowship.

\clearpage








\end{document}